\documentclass[journal,10pt]{IEEEtran}
\ifCLASSINFOpdf
   \usepackage[pdftex]{graphicx}
\else
\fi
\usepackage{amsmath}
\usepackage{array}
\usepackage{float}
\usepackage{amsmath}
\usepackage{amsfonts}
\usepackage{lscape}
\usepackage{mathtools}
\usepackage{amssymb}
\usepackage{float}
\usepackage{xcolor}
\usepackage{enumerate}
\usepackage{isomath}
\usepackage{bm}
\usepackage{cite}
\usepackage{caption}
\usepackage{subfig}
\usepackage{lipsum}
\usepackage{graphicx}
\usepackage{algorithm,algcompatible,amsmath}
\algnewcommand\INPUT{\item[\textbf{Input:}]}%
\algnewcommand\OUTPUT{\item[\textbf{Output:}]}%
\restylefloat{table}
\usepackage{tikz} 
\usepackage{lipsum}

\usetikzlibrary{arrows, positioning, automata}
\usetikzlibrary{shapes.geometric, arrows}
\def\xfoo#1^#2\relax#3\valign{%
\mathbf{#1}\ifx\valign#2\valign\else^{\mathbf{#2}}\fi}

\def \alphah{\hat{\alpha}}
\def \betah{\hat{\beta}}

\def\Ber{{\rm Bern}}

\def\Xbf{\mathbf{X}}
\def\Ybf{\mathbf{Y}}

\def \CLCI{\mathrm{CL/CI}}
\def \CLDI{\mathrm{CL/DI}}
\def \DLCI{\mathrm{DL/CI}}
\def \DLDI{\mathrm{DL/DI}}

\def \Nscr{\mathcal{N}}
\def \Wscr{\mathscr{W}}

\def \Xscrhat{{\widehat{\mathcal{X}}}}
\def \Xhatbf{\widehat{\mathbf{X}}}

\usepackage{amsmath, amssymb, bbm, xspace}
\usepackage{epsfig}
\usepackage{longtable}
\usepackage{color}
\usepackage{mathrsfs}
\usepackage{comment}

\usepackage{courier}

\newtheorem{theorem}{Theorem}[section]

\newtheorem{lemma}{Lemma}[section]

\newtheorem{proposition}[theorem]{Proposition}

\newtheorem{definition}{Definition}[section]

\def\bkE{{\rm I\kern-.17em E}}
\def\bk1{{\rm 1\kern-.17em l}}
\def\bkD{{\rm I\kern-.17em D}}
\def\bkR{{\rm I\kern-.17em R}}
\def\bkP{{\rm I\kern-.17em P}}

\def\bkZ{{\bf{Z}}}

\def\bkE{{\rm I\kern-.17em E}}
\def\bk1{{\rm 1\kern-.17em l}}
\def\bkD{{\rm I\kern-.17em D}}
\def\bkR{{\rm I\kern-.17em R}}
\def\bkP{{\rm I\kern-.17em P}}

\makeatletter
\newcommand{\pushright}[1]{\ifmeasuring@#1\else\omit\hfill$\displaystyle#1$\fi\ignorespaces}
\newcommand{\pushleft}[1]{\ifmeasuring@#1\else\omit$\displaystyle#1$\hfill\fi\ignorespaces}
\makeatother

\def\bkZ{{\bf{Z}}}
\def\b12{(\beta_1,\beta_2)}

\newcounter{example}
\renewcommand{\theexample}{\thesection.\arabic{example}}

\newcounter{remark}
\renewcommand{\theremark}{\thesection.\arabic{remark}}

\def\Xscr{\mathcal{X}}
\def\Yscr{\mathcal{Y}}

\def\Ebb{\mathbb{E}}
\newlength{\noteWidth}
\long\def\notes#1{\ifinner
{\tiny #1}
\else
\marginpar{\parbox[t]{\noteWidth}{\raggedright\tiny #1}}
\fi\typeout{#1}}

 \def\notes#1{\typeout{read notes: #1}} 

\newcommand{\ie}{i.e.\@\xspace} 
\newcommand{\eg}{e.g.\@\xspace} 
\newcommand{\etal}{et al.\@\xspace} 

\def\Ebb{\mathbb{E}}

\def\Ibb{{\mathbb{I}}}

\def\spose#1{\hbox to 0pt{#1\hss}}

\def\text #1{\hbox{\quad#1\quad}}

\def\nthinsp{\mskip -2   mu}

\def\superstar{^{\raise 0.5pt\hbox{$\nthinsp *$}}}
\def\SUPERSTAR{^{\raise 0.5pt\hbox{$*$}}}

\def\lamstarT {\lambda^{\raise 0.5pt\hbox{$\nthinsp *$}T}}

\def\Fscr{{\cal F}}

\def\Mscr{{\cal M}}

\def\Pscr{{\cal P}}
\def\Qscr{{\cal Q}}

\def\Wscr{{\cal W}}
\def\Mscr{{\cal M}}
\def\Nscr{{\cal N}}
\def\Rscr{{\cal R}}

\def\Cscr{{\cal C}}

\def\Xscr{{\cal X}}
\def\Yscr{{\cal Y}}

\def\Xhat{\widehat X}

\def\non{\nonumber}

\let\forallnew\forall
\renewcommand{\forall}{\forallnew\ }
\let\forall\forallnew

		\def\bkE{{\rm I\kern-.17em E}}
		\def\bk1{{\rm 1\kern-.17em l}}
		\def\bkD{{\rm I\kern-.17em D}}
		\def\bkR{{\rm I\kern-.17em R}}
		\def\bkP{{\rm I\kern-.17em P}}
		\def\bkY{{\bf \kern-.17em Y}}
		\def\bkZ{{\bf \kern-.17em Z}}
		\def\bkC{{\bf  \kern-.17em C}}

{\begin{list}{}%
         {\setlength{\leftmargin}{#1}}%
         \item[]%
}
{\end{list}}

		\def\bsp{\begin{split}}
		\def\beq{\begin{eqnarray}}
		\def\bal{\begin{align*}}
		\def\bc{\begin{center}}
		\def\be{\begin{enumerate}}
		\def\bi{\begin{itemize}}
		\def\bs{\begin{small}}
		\def\bS{\begin{slide}}
		\def\ec{\end{center}}
		\def\ee{\end{enumerate}}
		\def\ei{\end{itemize}}
		\def\es{\end{small}}
		\def\eS{\end{slide}}
		\def\eeq{\end{eqnarray}}
		\def\eal{\end{align*}}
		\def\esp{\end{split}}
		\def\qed{ \vrule height7.5pt width7.5pt depth0pt}  

	\def\cp2problem#1#2#3#4{\fbox
		 {\begin{tabular*}{0.9\textwidth}
			{@{}l@{\extracolsep{\fill}}l@{\extracolsep{6pt}}l@{\extracolsep{\fill}}c@{}}
				#1 & & $#4 $ 
			\end{tabular*}}}

		\def\bkE{{\rm I\kern-.17em E}}
		\def\bk1{{\rm 1\kern-.17em l}}
		\def\bkD{{\rm I\kern-.17em D}}
		\def\bkR{{\rm I\kern-.17em R}}
		\def\bkP{{\rm I\kern-.17em P}}
		
		\def\bkZ{{\bf{Z}}}

\newcommand {\beeq}[1]{\begin{equation}\label{#1}}
\newcommand {\eeeq}{\end{equation}}
\newcommand {\bea}{\begin{eqnarray}}
\newcommand {\eea}{\end{eqnarray}}

\def\texitem#1{\par\smallskip\noindent\hangindent 25pt
               \hbox to 25pt {\hss #1 ~}\ignorespaces}

\def\bsp{\begin{split}}
		\def\beq{\begin{eqnarray}}
		\def\bal{\begin{align*}}
		\def\bc{\begin{center}}
		\def\be{\begin{enumerate}}
		\def\bi{\begin{itemize}}
		\def\bs{\begin{small}}
		\def\bS{\begin{slide}}
		\def\ec{\end{center}}
		\def\ee{\end{enumerate}}
		\def\ei{\end{itemize}}
		\def\es{\end{small}}
		\def\eS{\end{slide}}
		\def\eeq{\end{eqnarray}}
		\def\eal{\end{align*}}
		\def\esp{\end{split}}
		\def\qed{ \vrule height7.5pt width7.5pt depth0pt}  

\usepackage{amsmath, amssymb, xspace}
\usepackage{epsfig}
\usepackage{longtable}
\usepackage{color}
\usepackage{mathrsfs}
\usepackage{subfig}
\def\Cscr{{\cal C}}
\def\Nscr{{\cal N}}

\begin{document}
%
\title{An Information-Theoretic Analysis of The Cost of Decentralization for Learning and Inference Under Privacy Constraints}
%
%
%

\author{Sharu~Theresa~Jose,~\IEEEmembership{Member,~IEEE,}
        and~Osvaldo~Simeone,~\IEEEmembership{Fellow,~IEEE,}
\thanks{The authors are with the Department of Engineering of King’s College London, UK (emails: sharu.jose@kcl.ac.uk, osvaldo.simeone@kcl.ac.uk).
They have received funding from the European Research Council
(ERC) under the European Union’s Horizon 2020 Research and Innovation
Programme (Grant Agreement No. 725731).}
}

\maketitle

\begin{abstract}
In vertical federated learning (FL), the features of a data sample are distributed across multiple agents. As such, inter-agent collaboration can be beneficial not only during the learning phase, as is the case for standard horizontal FL, but also during the inference phase. A fundamental theoretical question in this setting is how to quantify the cost, or performance loss, of decentralization for learning and/or inference. 
 In this paper, we consider general supervised learning problems with any number of agents, and provide
a novel information-theoretic quantification of the cost of decentralization in the presence of privacy constraints on inter-agent communication within a Bayesian framework. The cost of decentralization for learning and/or inference is shown to be quantified in terms of conditional mutual information terms involving features and label variables.

\end{abstract}


%
\IEEEpeerreviewmaketitle

\section{Introduction}
Consider a vertical federated learning (FL) framework  in which attributes (or features) of a data sample are distributed across multiple local agents. 
Such scenarios arise in many practical settings, including sensor arrays with component sensors being geographically distributed \cite{li2020federated}; or  insurance companies utilizing health records from different hospitals \cite{vepakomma2018split}. Note that vertical FL is distinct from horizontal FL, which is studied in the overwhelming majority of papers on the subject  \cite{li2020federated}: In horizontal FL, agents have independent data points, while in vertical FL the agents share the same data points, whose features are partitioned across agents. As explained in \cite{verma2019federated}, \cite{chencollaboration}, in vertical FL  settings, the inter-agent collaboration can be beneficial not only  during the learning phase, as for horizontal FL, but also during the inference phase. It is then important to understand at a fundamental theoretical level if decentralization, wherein agents use only local data for learning and/or inference, entails a significant performance loss as compared to collaborative learning and/or inference. This is the subject of this paper.

\begin{figure}[h!]
    \centering
     \includegraphics[scale=0.32,trim=0.5in 0.1in 0in 0in,clip=true]{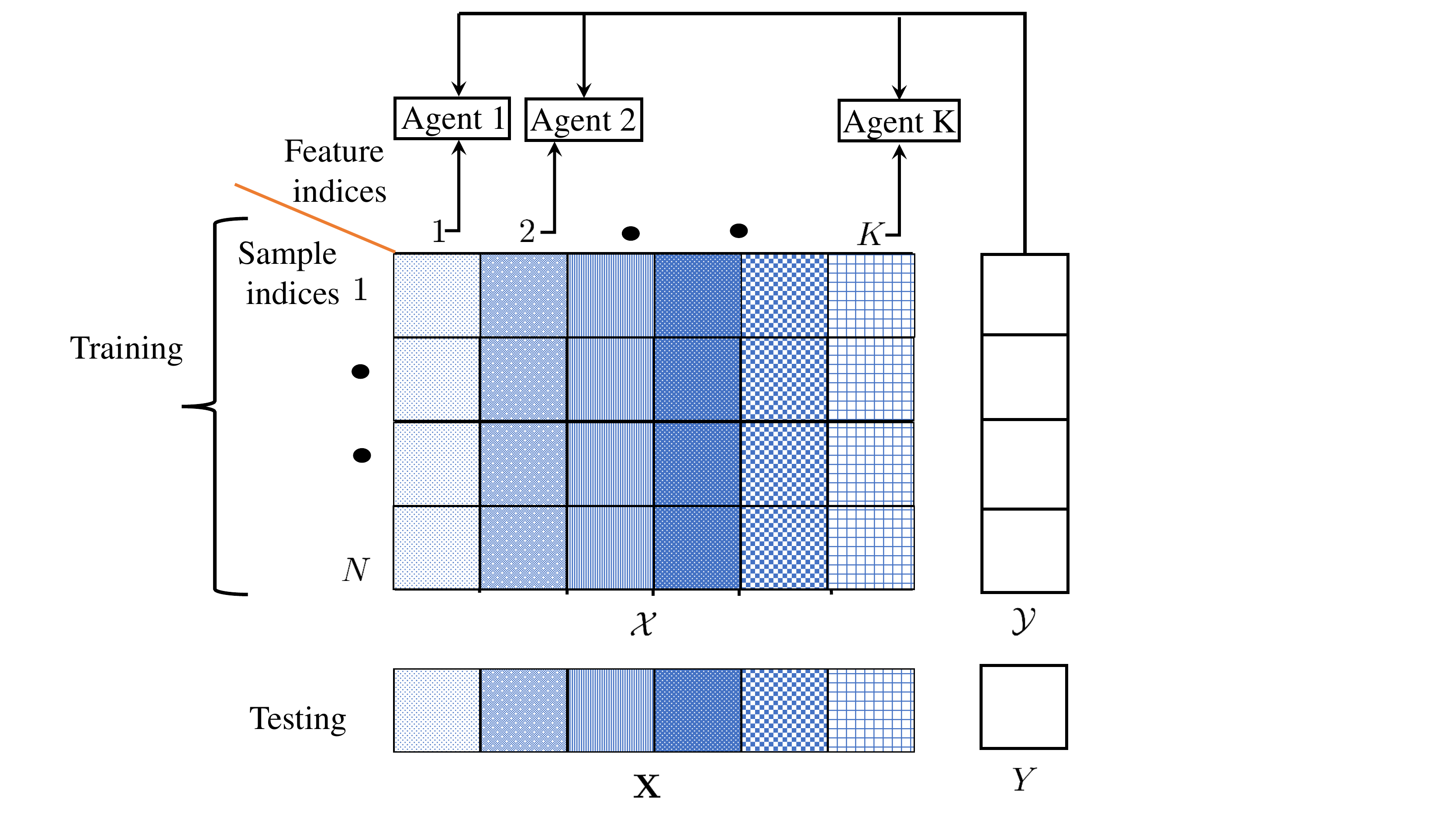}
    \caption{Illustration of the vertical federated learning (FL) setup under study.}
    \label{fig:verticalFL}
    \vspace{-0.6cm}
\end{figure} 
As a first attempt in this direction, Chen \etal \cite{chencollaboration} address this problem by studying a binary classification problem in which each class corresponds to a bivariate Gaussian distribution over two input features, which are vertically distributed between two agents. The authors identify four collaboration settings depending on whether collaboration is done during learning and/or inference phases as collaborative learning-collaborative inference (CL/CI), collaborative learning-decentralized inference (CL/DI), decentralized learning-collaborative inference (DL/CI), and decentralized learning-decentralized inference (DL/DI). By taking a frequentist approach, the authors compare the classification error rates achieved under these four settings.

In this work, inspired by \cite{chencollaboration}, we develop a novel \textit{information-theoretic} approach to quantify the cost of decentralization for \textit{general} supervised learning problems with \textit{any} number of agents and under \textit{privacy} constraints. Specifically, we consider a supervised learning problem defined by an arbitrary joint distribution $P_{\Xbf,Y|W}$ involving the feature vector $\Xbf$ and label $Y$, with the feature vector vertically partitioned between any number of local agents (see Figure~\ref{fig:verticalFL}). The agents are assumed to collaborate via a stochastic \textit{aggregation mechanism} that maps the local distributed features to a shared feature to be used by each agent during learning and/or inference. To limit the information leakage from the shared feature to an adversarial eavesdropper, unlike \cite{chencollaboration}, privacy constraints are imposed on the aggregation mapping. 
By adopting a Bayesian framework, we characterize the average predictive performance of the four settings -- CL/CI, CL/DI, DL/CI, and DL/DI -- under privacy constraints via  information-theoretic metrics.  Finally, we illustrate the relation between the four collaboration settings with/without privacy constraints on two numerical examples.

\section{Problem Formulation}\label{sec:problem_formulation}
\noindent{\textbf{{Setting}:}}
We study a vertical federated learning (FL) setting with $K$ agents that can cooperate during the learning and/or inference phases of operation of the sytem. Our main goal is to quantify using information-theoretic metrics the benefits of cooperation for learning and/or inference.
 We focus on a supervised learning problem, in which each  data point corresponds to a tuple $(\Xbf,Y)$ encompassing the $K$-dimensional feature vector $\Xbf=(X_1,\hdots,X_K)$ and 
 the scalar output label $Y$. As illustrated in Figure~\ref{fig:verticalFL}, each $k$th feature $X_k$ in vector $\Xbf$ is observed only at the $k$th agent, while the output label $Y$ is observed at all the $K$ agents \cite{liu2019communication,gu2020privacy}. Features and labels can take values in arbitrary alphabets. The unknown data distribution is assumed to belong to a model class $\{P_{\Xbf,Y|W}:W \in \Wscr\}$ of joint distributions that are identified by a model parameter vector $W$ taking values in some space $\Wscr$. Adopting a Bayesian approach, we endow the model parameter vector with a prior distribution $P_W$. 
 
 As illustrated in Figure~\ref{fig:verticalFL}, let $(\Xscr,\Yscr)=\{(\Xbf_1,Y_1),\hdots, (\Xbf_N,Y_N)\}$ denote a training data set of $N$ labelled samples which, when conditioned on model parameter $W$, are assumed to be generated i.i.d. according to distribution $P_{\Xbf,Y|W}$.  The $N \times K$ matrix $\Xscr $ collects the $K$-dimensional feature vectors $\{\Xbf_n\}_{n=1}^N$ by rows. We denote as  $X_{n,k}$, the $(n,k)$th element of matrix $\Xscr$, for $n=1,\hdots,N,$ and $k=1,\hdots,K$; and
as $\Xscr_{k}=[X_{1,k},\hdots,X_{N,k}]^T$ ($[\cdot]^T$ is the transpose operation) the $k$th column of the data matrix, which corresponds to the observations of agent $k$. The goal of the system is to use the training data set $(\Xscr,\Yscr)$ to enable the agents to predict the label of a new, previously unseen, test feature input $\Xbf$.
The joint distribution of model parameter $W$,  training data $(\Xscr,\Yscr)$ and test data $(\Xbf,Y)$ can be written as
\begin{align}
    P_{W,\Xscr,\Yscr,\Xbf,\Ybf}=P_W \otimes \underbrace{(P_{X_1,\hdots,X_K,\Ybf|W})^{\otimes N}}_{\mbox{training}} \otimes \underbrace{P_{X_1,\hdots,X_K,Y|W}}_{\mbox{testing}}, \label{eq:jointdistribution}
\end{align}
with $\otimes$ representing the product of distributions and $(\cdot)^{\otimes N}$ denoting the corresponding $N$-fold product.

\noindent{\textbf{{Collaborative/decentralized learning/inference}}:}
In the learning phase, training data is used to infer the model parameter $W$, enabling the agents in the inference phase to make predictions about test label $Y$ given the test feature vector $\Xbf$ based on the model $P_{\Xbf,Y|W}$. Either or both learning and inference phases can be carried out collaboratively by the agents or in a decentralized fashion, \ie, separately by each agent. When collaborating for learning or inference, the $K$ agents share their locally observed feature data via a third party, \eg, an edge server in a wireless cellular architecture. The operation of the third party is modelled as a stochastic aggregation mapping $P_{\Xhatbf|X_1,\hdots,X_k}=P_{\Xhatbf|\Xbf}$ from the input $K$ local features to an output shared feature $\Xhatbf$. As detailed next, for learning, the mapping $P_{\Xhatbf|\Xbf}$ is applied independently to each data point. Furthermore, as we also detail later in this section, we impose privacy constraints on the aggregation mapping $P_{\Xhatbf|\Xbf}$ so that the shared feature $\Xhatbf$ does not reveal too much information about the local agents' features.

 We specifically distinguish the following four settings:\\
\noindent $\bullet$ \textit{Collaborative learning-collaborative inference} (CL/CI): Agents collaborate during both learning and inference phases by sharing information about their respective features. Accordingly, during learning, each agent has access to the shared training data features $\widehat{\Xscr}=(\Xhatbf_1,\hdots, \Xhatbf_N)$ with $\Xhatbf_n \sim P_{\Xhatbf|\Xbf=\Xbf_n}$ being generated independently by the third party, in addition to its own observed local feature data $\Xscr_k$. Furthermore, during inference, agent $k$ can use the shared test feature $\Xhatbf \sim P_{\Xhatbf|\Xbf=\Xbf}$ in addition to its own observation $X_k$ in order to predict the test label $Y$.\\
   \noindent $\bullet$ \textit{Collaborative learning-decentralized inference} (CL/DI): Agents  collaborate only during learning by sharing information about their respective features as explained above, while inference is decentralized. Accordingly, during inference, each $k$th agent uses the $k$th feature $X_k$ of test feature vector $\Xbf$ in order to predict the test label $Y$.\\
  \noindent $\bullet$ \textit{Decentralized learning-collaborative inference} (DL/CI): Agents collaborate for inference, while each $k$th agent is allowed to use only its observed training data $(\Xscr_{k},\Yscr)$ during the learning phase.\\
   \noindent $\bullet$ \textit{Decentralized learning-decentralized inference} (DL/DI): Agents operate independently, with no cooperation in either learning or inference phases.

\noindent {\textbf{Privacy constraints}}: The aggregation mapping $P_{\Xhatbf|\Xbf}$ shares the output feature $\Xhatbf$ with each of the $K$ local agents during collaborative learning and/or inference. To account for privacy constraints concerning agents' data, we limit the amount of information that a ``curious" eavesdropper  may be able to obtain about the local features' data from observing $\Xhatbf$. To this end, we impose the following privacy constraint on the aggregation mapping so that the shared feature $\Xhatbf$ does not leak too much information about the local features $X_k$ of all agents $k=1,\hdots,K$.

The aggregation mapping $P_{\Xhatbf|\Xbf}$ is said to be $\epsilon$-private if
\begin{align}
I(\Xhatbf;X_k|X^{(-k)}) \leq \epsilon, \quad \mbox{for all} \hspace{0.2cm} k=1,\hdots,K, \label{eq:epsilon-privacy-def}
\end{align}
where $X^{(-k)}=(X_1,\hdots,X_{k-1},X_{k+1},\hdots,X_K)$.
The constraint \eqref{eq:epsilon-privacy-def} measures privacy against a strong eavesdropper that knows all features except the $k$th feature $X_k$. Specifically, the conditional mutual information $I(\Xhatbf;X_k|X^{(-k)})$ quantifies the additional information about $X_k$ gained by the eavesdropper upon observing the shared feature $\Xhatbf$. As such, the metric is also relevant as a privacy measure against ``curious" agents.

We note that although the privacy constraint in \eqref{eq:epsilon-privacy-def} bears resemblance to the MI-differential privacy (MI-DP) constraint introduced in \cite{cuff2016differential}, the condition \eqref{eq:epsilon-privacy-def} does not have the same operational meaning. In fact, the MI-DP constraint in \cite{cuff2016differential},\cite{yagli2020information} ensures differential privacy for individual i.i.d. data samples of a training data set, and it relies on a mechanism that applies on the entire data set during learning. In contrast, the constraint \eqref{eq:epsilon-privacy-def} accounts for the privacy of correlated local features via a per-sample masking mechanism, and it applies to both learning and inference phases.

\noindent \textbf{Predictive loss under privacy constraints}: In all the four settings described above,
any agent $k$ uses the available training data $(\widetilde{\Xscr}_k,\Yscr)$, with $\widetilde{\Xscr}_k$ being equal to $\Xscr_k$ for decentralized learning and to $(\Xscr_k,\widehat{\Xscr})$ for collaborative learning, in order to infer the model parameter $W$. The inferred model is then used to predict the label $Y$ given the test feature input $\widetilde{\Xbf}_k$, with $\widetilde{\Xbf}_k$ being equal to $X_k$ for decentralized inference and to $(X_k,\Xhatbf)$ for collaborative learning. We impose that the aggregation mapping $P_{\Xhatbf|\Xbf}$ must satisfy the privacy constraint in \eqref{eq:epsilon-privacy-def}. 

 The joint operation of learning and inference at agent $k$ can be accordingly described via a stochastic predictive distribution $Q_{Y|\widetilde{\Xscr}_k,\Yscr,\widetilde{\Xbf}_k}$  on the test label $Y$ given the training data $(\widetilde{\Xscr}_k,\Yscr)$ and test feature input $\widetilde{\Xbf}_k$. Note that this stochastic mapping can account for arbitrary choices of learning and inference algorithms. By optimizing over aggregation mapping  as well as over learning and inference algorithms,  we define the $\epsilon$-private  predictive loss as
\begin{align}
    \hspace{-0.1cm}\Rscr(\epsilon)&=\hspace{-0.3cm}\min_{\substack{P_{\Xhatbf|\Xbf} \\\in \Pscr(\Xhatbf|\Xbf)}} \hspace{-0.1cm}\hspace{-0.1cm}\max_{k=1,\hdots,K} \hspace{-0.2cm}\min_{\substack{Q_{Y|\widetilde{\Xscr}_k,\Yscr,\widetilde{\Xbf}_k} \\ \in \mathcal{Q}(Y|\widetilde{\Xscr}_k,\Yscr,\widetilde{\Xbf}_k)}}\hspace{-0.5cm} \Ebb_{P_{Y,\widetilde{\Xscr}_k,\Yscr,\widetilde{\Xbf}_k}}\hspace{-0.1cm}\Bigl[-\log Q_{Y|\widetilde{\Xscr}_k,\Yscr,\widetilde{\Xbf}_k} \Bigr] \non \\
   & \quad \mbox{s.t}  \hspace{0.2cm}  I(\Xhatbf;X_k|X^{(-k)}) \leq \epsilon \quad \mbox{for all} \hspace{0.2cm} k=1,\hdots,K.
    \label{eq:R_epsilon}
\end{align} In \eqref{eq:R_epsilon}, the aggregation  mapping $P_{\Xhatbf|\Xbf}$ is optimized over some specified family $\Pscr(\Xhatbf|\Xbf)$ of conditional distributions $P_{\Xhatbf|\Xbf}$ in order to minimize the worst case predictive loss across the agents under constraint \eqref{eq:epsilon-privacy-def}. Furthermore, the inner optimization is over a class of predictive distributions $\mathcal{Q}(Y|\widetilde{\Xscr}_k,\Yscr,\widetilde{\Xbf}_k) $.

In the absence of privacy constraints, \ie, when $\epsilon=\infty$, assuming that the distribution family $\Pscr(\Xhatbf|\Xbf)$ is sufficiently large, the optimal aggregation mapping $P_{\Xhatbf|\Xbf}$ puts its entire mass on the output shared feature $\Xhatbf=\Xbf$. As such, under collaborative learning, each agent $k$ uses the entire feature data \ie, $\widetilde{\Xscr}_k=\Xscr$;  and, under collaborative inference, it uses the entire test feature vector $\widetilde{\Xbf}_k=\Xbf$.  The predictive loss \eqref{eq:R_epsilon} in this case evaluates as
\begin{align}
\Rscr(\infty)=\max_{k=1,\hdots,K} \hspace{-0.2cm}\min_{\substack{Q_{Y|\widetilde{\Xscr}_k,\Yscr,\widetilde{\Xbf}_k}\\ \in \mathcal{Q}(Y|\widetilde{\Xscr}_k,\Yscr,\widetilde{\Xbf}_k ) }} \hspace{-0.3cm}\Ebb_{P_{Y,\widetilde{\Xscr}_k,\Yscr,\widetilde{\Xbf}_k}}\Bigl[-\log Q_{Y|\widetilde{\Xscr}_k,\Yscr,\widetilde{\Xbf}_k} \Bigr]. \label{eq:Bayesianrisk_generaldefinition}
\end{align} 
The predictive loss \eqref{eq:Bayesianrisk_generaldefinition} represents the worst-case minimum average cross-entropy loss across all agents, that can be obtained given the information about the training data set and the test input feature \cite{xu2020minimum}.

\section{ Preliminaries and Fully Collaborative Benchmark}\label{sec:collaborativeLI}
In this section, we first provide a brief explanation of the main information-theoretic metrics used in this work. Then, we define and derive the average predictive loss for the benchmark case in which both learning and inference are collaborative.

\noindent \textbf{{Information-theoretic Metrics}}:
Let $A$ and $B$ denote two (discrete or continuous) random variables with joint distribution $P_{A,B}$, and with corresponding marginals $P_A$ and $P_B$. The joint entropy of $A$ and $B$, denoted $H(A,B)$, is defined as $H(A,B)=\Ebb_{P_{A,B}}[-\log P_{A,B}]$, with $\Ebb_{P}[\cdot]$ denoting the expectation with respect to distribution $P$. More generally, the conditional entropy of $A$ given $B$ is defined as $H(A|B)=\Ebb_{P_{A,B}}[-\log P_{A|B}]$, where $P_{A|B}=P_{A,B}/P_B$ is the conditional distribution of $A$ given $B$. By the chain rule, we have the relationship $H(A,B)=H(B)+H(A|B)$; and we also have the property that conditioning  reduces entropy \cite{cover06elements} \ie,
$
H(A|B)\leq H(A) $
The mutual information $I(A;B)$ between the random variables is defined as $
I(A;B)=\Ebb_{P_{A,B}}\biggl[\log \biggl(\frac{P_{A,B}}{P_AP_B}\biggr)\biggr]. $ Finally, for random variables $A,B$ and $C$ with joint distribution $P_{A,B,C}$,  the conditional mutual information $I(A;B|C)$ between $A$ and $B$ given $C$ is defined as $I(A;B|C)=\Ebb_{P_{A,B,C}}\biggl[ \log \biggl(\frac{P_{A,B|C}}{P_{A|C}P_{B|C}}\biggr)\biggr]$.
\\

\noindent \textbf{Private collaborative learning-collaborative inference} (CL/CI): As a benchmark, we now study the predictive loss \eqref{eq:R_epsilon} 
for the CL/CI setting. The $\epsilon$-private predictive loss \eqref{eq:R_epsilon} of CL/CI is given as
\begin{align}
    & \Rscr^{\CLCI}(\epsilon)\non \\&=\hspace{-0.3cm}\min_{\substack{P_{\Xhatbf|\Xbf} \\\in \Fscr(\Xhatbf|\Xbf)}}\hspace{-0.3cm} \max_{k=1,\hdots,K}\hspace{-0.3cm}\min_{\substack{Q_{Y|\widehat{\Xscr},\Xscr_k,\Yscr,\Xhatbf,X_k}\\ \in \Qscr(Y|\widehat{\Xscr},\Xscr_k,\Yscr,\Xhatbf,X_k )}}\hspace{-0.8cm} \Ebb_{P_{Y,\widehat{\Xscr},\Xscr_k,\Yscr,\Xhatbf,X_k}}\hspace{-0.1cm}\Bigl[\hspace{-0.05cm}-\log Q_{Y|\widehat{\Xscr},\Xscr_k,\Yscr,\Xhatbf,X_k} \Bigr] \label{eq:CLCIrisk_privacy1}\\
    &=\min_{\substack{P_{\Xhatbf|\Xbf} \\\in \Fscr(\Xhatbf|\Xbf)}} \max_{k=1,\hdots,K} H(Y|\widehat{\Xscr},\Xscr_k,\Yscr,\Xhatbf,X_k), \label{eq:CLCIrisk_privacy2}
\end{align}where \begin{align}
\Fscr(\Xhatbf|\Xbf)=\{P_{\Xhatbf|\Xbf} \in \Pscr(\Xhatbf|\Xbf): \mbox{constraint \eqref{eq:epsilon-privacy-def} holds}\} \label{eq:feasiblespace} 
\end{align} is the feasible space of conditional distributions satisfying the privacy constraint \eqref{eq:epsilon-privacy-def}. The equality in \eqref{eq:CLCIrisk_privacy2} holds under the assumption that the distribution family $\Qscr(Y|\widehat{\Xscr},\Xscr_k,\Yscr,\Xhatbf,X_k )$ is sufficiently large to include the posterior distribution $P_{Y|\widehat{\Xscr},\Xscr_k,\Yscr,\Xhatbf,X_k}$. In fact, for any  fixed aggregation mapping $P_{\Xhatbf|\Xbf} \in \Fscr(\Xhatbf|\Xbf)$, the posterior minimizes the cross-entropy metric in \eqref{eq:CLCIrisk_privacy1}.
 In a similar manner, when no privacy constraints are imposed, \ie, when $\epsilon=\infty$, and the family $\Pscr(\Xhatbf|\Xbf)$ is large enough, the Bayesian predictive loss  \eqref{eq:Bayesianrisk_generaldefinition} can be exactly characterized as
\begin{align}
\Rscr^{\CLCI}(\infty)=H(Y|\Xbf,\Xscr,\Yscr).
\end{align} 

\section{ Cost of Decentralization Under Privacy Constraints}\label{sec:information_theoreticcost}
In this section, we use the benchmark predictive loss \eqref{eq:CLCIrisk_privacy2} observed under the ideal CL/CI setting to evaluate the cost of decentralization in the learning and/or inference phases under privacy constraints.

\begin{lemma}\label{lem:Bayesianrisk_schemes}
The $\epsilon$-private predictive losses of decentralized learning and/or inference are given as
\begin{align}
 \Rscr^{\CLDI}(\epsilon)&=\hspace{-0.2cm}\min_{P_{\Xhatbf|\Xbf} \in \Fscr(\Xhatbf|\Xbf)} \max_{k=1,\hdots,K} H(Y|X_k,\Xscr_k,\widehat{\Xscr},\Yscr)\label{eq:risk_2}\\
\Rscr^{\DLCI}(\epsilon)&=\hspace{-0.2cm}\min_{P_{\Xhatbf|\Xbf} \in \Fscr(\Xhatbf|\Xbf)}  \max_{k=1,\hdots,K} H(Y|X_k,\Xhatbf,\Xscr_{k},\Yscr)\label{eq:risk_3}\\
 \Rscr^{\DLDI}(\epsilon)&= \max_{k=1,\hdots,K} H(Y|X_k,\Xscr_{k},\Yscr)\label{eq:risk_4},
\end{align}
where set $\Fscr(\Xhatbf|\Xbf)$ is as defined in \eqref{eq:feasiblespace}.
\end{lemma}

Note that the predictive loss \eqref{eq:risk_4} of the fully decentralized DL/DI setting does not depend on the privacy parameter $\epsilon$ since decentralization does not entail any privacy loss. Therefore, in the absence of privacy constraints, we have $\Rscr^{\DLDI}(\infty)=\Rscr^{\DLDI}(\epsilon)$, while the predictive losses in \eqref{eq:risk_2}--\eqref{eq:risk_3} evaluate as
\begin{align}
\Rscr^{\CLDI}(\infty)&= \max_{k=1,\hdots,K}H(Y|X_k,\Xscr,\Yscr), 
\end{align}
\begin{align} \Rscr^{\DLCI}_k(\infty)&=\max_{k=1,\hdots,K} H(Y|\Xbf,\Xscr_k,\Yscr),
\end{align}under the assumption of sufficiently large $\Pscr(\Xhatbf|\Xbf)$.
Furthermore, using the property  that conditioning reduces entropy results in the following relation between the predictive losses of the four schemes -- CL/CI, CL/DI, DL/CI and DL/DI -- in the absence of privacy constraints,
\begin{align}
    \Rscr^{\CLCI}(\infty) &\leq \min\{\Rscr^{\CLDI}(\infty), \Rscr^{\DLCI}(\infty)\} \non\\&\leq \max\{\Rscr^{\CLDI}(\infty), \Rscr^{\DLCI}(\infty)\} \non \\& \leq \Rscr^{\DLDI}(\epsilon) \label{eq:relation_1}.
\end{align}

The difference between the $\epsilon$-private predictive risks of the decentralized and collaborative schemes capture the \textit{cost of decentralization}. Specifically, given two schemes $a,b \in \{$CL/CI, CL/DI, DL/CI, DL/DI$\}$ such that $\Rscr^a(\epsilon) \geq \Rscr^b (\epsilon)$, we define the cost of $a$ with respect to $b$ as
\begin{align}
\Cscr^{a-b}(\epsilon)=\Rscr^a(\epsilon)-\Rscr^b(\epsilon). \label{eq:cost_definition}
\end{align}
In the absence of privacy constraints $(\epsilon=\infty)$ and assuming symmetric agents so that the maximum in \eqref{eq:Bayesianrisk_generaldefinition} is attained for any $k=1,\hdots,K$, the cost of decentralization can be exactly characterized as in the following result.

\begin{proposition}
The cost of decentralization \eqref{eq:cost_definition} for $\epsilon=\infty$  and symmetric agents can be characterized for the $k$th learning agent as detailed in Table~\ref{table:1},
where $X^{(-k)}=(X_1,\hdots,X_{k-1},X_{k+1},\hdots,X_K)$ and $\Xscr^{(-k)}=(\Xscr_{1},\hdots,\Xscr_{k-1},\Xscr_{k+1},\hdots,\Xscr_{K})$.
\end{proposition}

The results in Table~\ref{table:1} have intuitive interpretations. For instance, the cost $\Cscr^{\CLDI-\CLCI}(\infty)=I(Y;X^{(-k)}|X_k,\Xscr,\Yscr)$ corresponds to the additional information about label $Y$ that can be obtained from observing the features $X^{(-k)}$ of other agents, given $\Xscr,\Yscr$ and $X_k$. Examples will be provided in the next section in which the cost of decentralization is evaluated also in the presence of privacy constraints based on \eqref{eq:CLCIrisk_privacy2}, \eqref{eq:risk_2}--\eqref{eq:risk_4}.


\begin{table*}[h!]
\caption{Cost of decentralization $\Cscr^{a-b}(\infty)$ ($a$ defines the column and $b$ the row)} \label{table:1}
\centering
\begin{tabular}{|c|c|c|c|c|}
\hline
 & CL/CI & CL/DI & DL/CI & DL/DI\\
\hline 
CL/CI & -- &  $I(Y;X^{(-k)}|X_k,\Xscr,\Yscr)$ &  $I(Y;\Xscr^{(-k)}|\Xbf,\Xscr_{k},\Yscr)$ & $I(Y;X^{(-k)},\Xscr^{(-k)}|X_k,\Xscr_{k},\Yscr)$ \\
& &  &  &  \\
\hline
CL/DI & -- & -- & &  $I(Y;\Xscr^{(-k)}|\Xscr_{k},X_k,\Yscr)$  \\
\hline
DL/CI & -- &  & -- & $I(Y;X^{(-k)}|X_k,\Xscr_{k},\Yscr)$\\
\hline
DL/DI &-- & -- & -- & --\\
\hline
\end{tabular}
\end{table*}

\section{Examples}
In this section, we consider two simple numerical examples to illustrate the cost of decentralization for learning and/or inference with and without the privacy constraints that was quantified in Section~\ref{sec:information_theoreticcost} for general models. 
\subsection{Two-Agent Non-Private Collaborative Learning and/or Inference} \label{sec:example1}
Consider two agents ($K=2$) observing binary joint features 
$X_1,X_2 \in \{0,1\}$, which have the joint distribution defined by the probability $r$ of the two features $X_1$ and $X_2$ being equal, \ie, 
$
\mathrm{Pr}[X_1=X_2]=r/2, \hspace{0.1cm} \mbox{with} \hspace{0.1cm} \mathrm{Pr}[X_1=1]= \mathrm{Pr}[X_2=1]=0.5. $
 Parameter $r$ quantifies the statistical dependencies between features $X_1$ and $X_2$ through the MI $I(X_1;X_2)=\log 2-H_b(r)$, where $H_b(r)=-r \log(r) -(1-r)\log(1-r)$ denotes the binary entropy with parameter $r$. Note that the MI takes the maximum value of $I(X_1;X_2)=1$ when $r=0$ or $1$, and the minimum value of $I(X_1;X_2)=0$ when $r=0.5$.  The output binary label $Y \in \{0,1\}$ depends on the feature vector $\Xbf$ through the model  \begin{align}
P_{Y=1|X_1,X_2,W}=\begin{cases} W_1 & \mbox{if}\hspace{0.2cm} X_1 \oplus X_2=0\\ W_2 &\mbox{if} \hspace{0.2cm} X_1 \oplus X_2=1 \end{cases}, \label{eq:datamodel_example}
\end{align} with model parameters $W=(W_1,W_2)$, where $\{W_1,W_2\} \in [0,1]$. Accordingly, $W_1$ and $W_2$ are the probabilities of the event $Y=1$ when $X_1$ and $X_2$ are equal or different, respectively. 
 We assume that the model parameters are a priori independent and distributed according to beta distributions as 
\begin{align}P_{W_1,W_2}={\rm Beta}(W_1|\alpha_1,\beta_1) {\rm Beta}(W_2|\alpha_2,\beta_2), \label{eq:prior_distribution} \end{align}
where  $\alpha_1,\beta_1,\alpha_2,\beta_2>0$ are fixed hyperparameters.
 \begin{figure}[h!]
    \centering
     \includegraphics[scale=0.4,trim=2in 1.4in 1.2in 0.7in,clip=true]{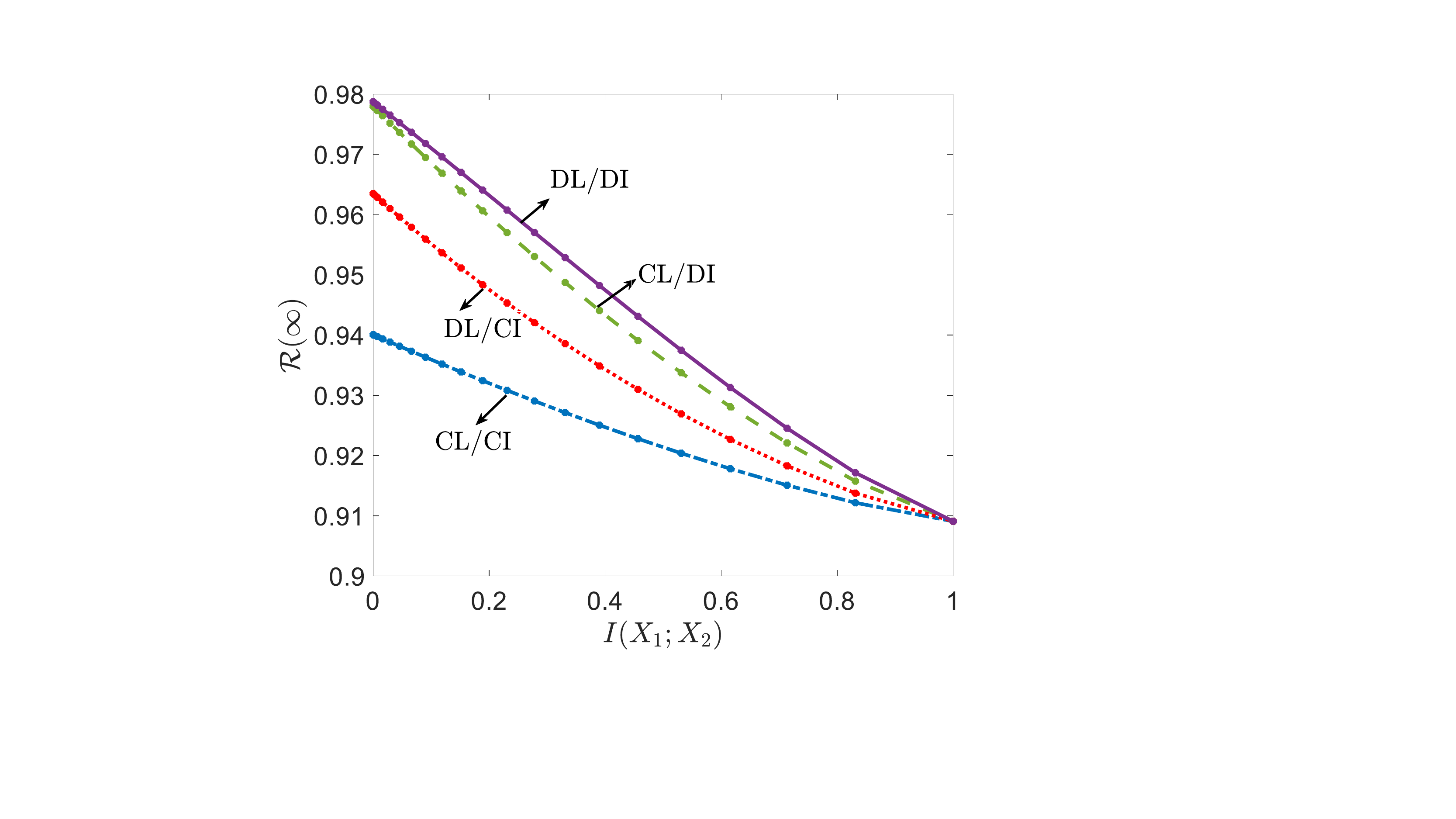}
    \caption{Predictive losses \eqref{eq:CLCIrisk_privacy2}, \eqref{eq:risk_2}--\eqref{eq:risk_4} for the four schemes under no privacy constraints ($\epsilon=\infty)$ as a function of the mutual information $I(X_1;X_2)$. ($\alpha_1=2$, $\beta_1=1.5$, $\alpha_2=1.5$, $\beta_2=2$ and $N=3$.) }
    \label{fig:1}
\end{figure}

Figure~\ref{fig:1} compares the predictive loss derived in Lemma~\ref{lem:Bayesianrisk_schemes} with no privacy constraints $(\epsilon=\infty)$ under the four schemes -- CL/CI, CL/DI, DL/CI and DL/DI -- as a function of the mutual information $I(X_1;X_2)$ between the components of the bivariate feature vector. The number of data samples is $N=3$ and other hyperparameters are set to $\alpha_1=2$, $\beta_1=1.5$, $\alpha_2=1.5$, and $\beta_2=2$. 
When the MI $I(X_1;X_2)$ is large, the predictive risks under collaborative and decentralized schemes are similar, and the cost of decentralization is negligible. This is because a larger MI $I(X_1;X_2)$ implies that each local agent's feature $X_k$, for $k=1,2$, is highly informative about the local feature $X^{(-k)}$ of the other agent, and no significant additional information can be obtained via collaboration. This applies to both learning and inference phases. Conversely, when the MI is small, decentralization entails a significant cost. In this example, centralized inference is more effective than centralized learning due to the importance of having access to both $X_1$ and $X_2$ in order to infer $Y$ by \eqref{eq:datamodel_example}.
\subsection{Three-Agent Private Collaborative Learning and/or Inference}\label{sec:example2}
We now extend the example in Section~\ref{sec:example1} by considering three agents ($K=3)$ and by imposing privacy constraints during collaboration in the learning and inference phases. The feature vector $\Xbf=(X_1,X_2,X_3)$ consists of three binary features $X_k\in \{0,1\}$ for $k=1,2,3$, where $X_1$ and $X_2$ are distributed as in Section~\ref{sec:example1}, and we have $\mathrm{Pr}[X_3|X_1=x_1,X_2=x_2]=\mathrm{Pr}[X_3|X_2=x_2]$ with $\mathrm{Pr}[X_3 \neq X_2|X_2=x_2]=(1-r)$. 
Generalizing the previous example, the output binary label $Y \in \{0,1\}$ depends on the feature vector $\Xbf$ through the model \begin{align}
P_{Y=1|\Xbf,W}=\begin{cases} W_1 & \mbox{if}\hspace{0.2cm} X_1 \oplus X_2 \oplus X_3=0\\ W_2 &\mbox{if} \hspace{0.2cm} X_1 \oplus X_2 \oplus X_3=1 \end{cases},
\end{align} where model parameters have the same prior distribution \eqref{eq:prior_distribution}.
The aggregation mapping $P_{\Xhatbf|\Xbf}$ produces a binary random variable $\Xhat\in \{0,1\}$ as \begin{align}
\Xhat=X_1 \oplus X_2 \oplus X_3 \oplus \xi, \quad \mbox{with} \hspace{0.2cm} \xi \sim \Ber(s), \label{eq:binary_moduloMAC}
\end{align}where parameter $s\in [0,1]$ is selected so as to guarantee the privacy constraints in \eqref{eq:epsilon-privacy-def}, which can be written as
\begin{align*}
\epsilon &\geq I(\Xhat;X_3|X_1,X_2)=-H_b(s)+H_b(s(1-r)+r(1-s))\\
\epsilon & \geq I(\Xhat;X_1|X_2,X_3)=-H_b(s)+H_b(sr+(1-r)(1-s))\\
\epsilon &\geq I(\Xhat;X_2|X_1,X_3)=-H_b(s)+2r(1-r)\log(2) \non \\&\qquad \hspace{0.8cm} +((1-r)^2+r^2)H_b\biggl(\frac{(1-r)^2s+r^2(1-s)}{(1-r)^2+r^2} \biggr).
\end{align*} 
\vspace{-0.5cm}
 \begin{figure}[h!]
    \centering
     \includegraphics[scale=0.34,trim=2.2in 1in 1.2in 0.5in,clip=true]{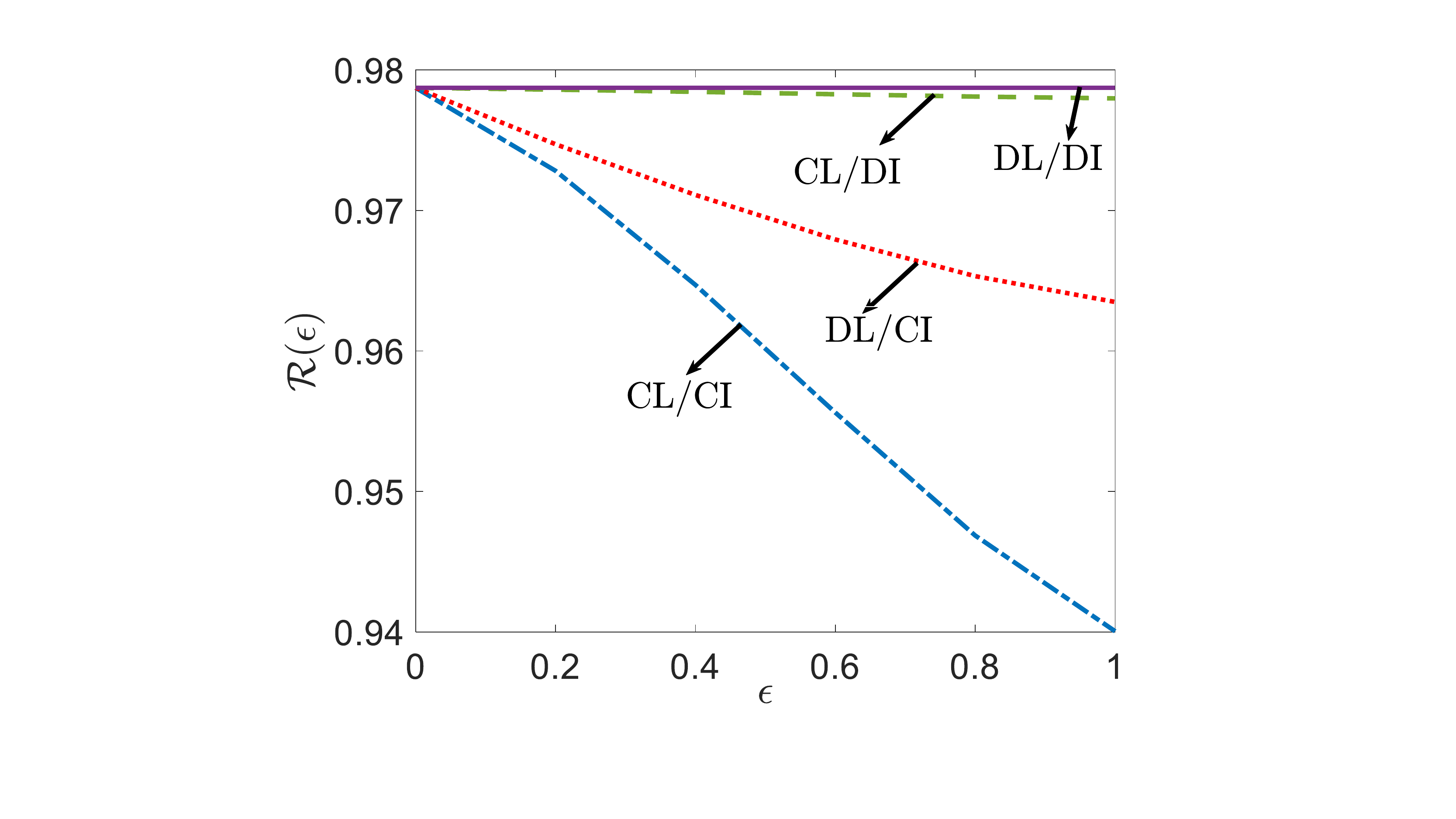}
    \caption{Predictive losses \eqref{eq:CLCIrisk_privacy2}, \eqref{eq:risk_2}--\eqref{eq:risk_4} for the four schemes as a function of privacy measure $\epsilon$. ($\alpha_1=2$, $\beta_1=1.5$, $\alpha_2=1.5$, $\beta_2=2$ and $N=3$.) }
    \label{fig:threeagents}
    \vspace{-0.2cm}
\end{figure}

Figure~\ref{fig:threeagents} compares the predictive loss $\Rscr(\epsilon)$ derived in Lemma~\ref{lem:Bayesianrisk_schemes} of the four schemes -- CL/CI, CL/DI, DL/CI and DL/DI -- as a function of the privacy parameter $\epsilon$ for fixed $r=0.5$. In the high-privacy regime, where $\epsilon$ is small, the shared feature $\Xhat$ is not informative about the local observed features, and collaborative learning/inference brings little benefit over the decentralized schemes. 
However, as $\epsilon$ increases, thereby  weakening privacy requirements, the shared feature $\Xhat$ becomes more informative about the observed feature vector $\Xbf$ and the cost of decentralization becomes increasingly significant, reaching its maximum value under no privacy, \ie, when $\epsilon=1$.
\vspace{-0.4cm}
\section{Conclusions}
This paper presents a novel information-theoretic characterization of the cost of decentralization during learning and/or inference in a vertical FL setting.  Under privacy constraints on the aggregation mechanism that enables inter-agent communications, we show, by adopting a Bayesian framework, that the average predictive performance of the four schemes can be quantified in terms of conditional entropies. Furthermore, when no privacy constraints are imposed, the cost of decentralization for symmetric agents is shown to be exactly characterized by conditional mutual information terms. Evaluating the derived metrics for real-world examples would generally require the implementation of mutual information estimators, and is left for future work.

\ifCLASSOPTIONcaptionsoff
  \newpage
\fi

\appendices
\vspace{-0.2cm}
\bibliographystyle{IEEEtran}
\bibliography{ref}

\end{document}